\documentclass[conference]{IEEEtran}
\IEEEoverridecommandlockouts

\usepackage{cite}
\usepackage{amsmath,amssymb,amsfonts}
\usepackage{algorithmic}
\usepackage{graphicx}
\usepackage{textcomp}
\usepackage{xcolor}
\usepackage{bm}
\usepackage{bbm}
\def\BibTeX{{\rm B\kern-.05em{\sc i\kern-.025em b}\kern-.08em
    T\kern-.1667em\lower.7ex\hbox{E}\kern-.125emX}}

\newtheorem{dfn}{Definition}
\newtheorem{thm}{Theorem}

\begin{document}

\title{The Conditional Regret-Capacity Theorem\\for Batch Universal Prediction}

\author{%
  \IEEEauthorblockN{Marco Bondaschi and Michael Gastpar}
  \IEEEauthorblockA{School of Computer and Communication Sciences\\
                    EPFL\\
                    Switzerland\\
                    Email: \{marco.bondaschi, michael.gastpar\}@epfl.ch}
}

\maketitle

\begin{abstract}
We derive a conditional version of the classical regret-capacity theorem. This result can be used in universal prediction to find lower bounds on the minimal batch regret, which is a recently introduced generalization of the average regret, when batches of training data are available to the predictor. As an example, we apply this result to the class of binary memoryless sources. Finally, we generalize the theorem to R\'{e}nyi information measures, revealing a deep connection between the conditional R\'{e}nyi divergence and the conditional Sibson's mutual information.
\end{abstract}


\section{Introduction}

Prediction of the continuation of a sequence from its own past is one of the central problems of statistics, science, and engineering.
One of its important roots is the influential work of Laplace~\cite{laplace1814essaiprob}.
In this paper, we are particularly interested in the study of universal prediction through the lens of information measures. This is a well-developed topic, see e.g. the survey paper~\cite{merhav1}, and has found many applications, such as compression \cite{ziv1, willems1}, gambling \cite{xie1} and machine learning \cite{fogel1,rosas1}. The cornerstone of this perspective is to consider prediction subject to logarithmic loss. Under this loss metric, performance criteria are expressed in terms of Kullback-Leibler divergence.

There is an important and fruitful connection between prediction and data compression. This is not counterintuitive: if a sequence can be successfully predicted, then one might expect that it can also be tightly compressed. Indeed, when logarithmic loss is considered, this connection can be made rigorous.
A landmark result is what is often referred to as the {\it redundancy-capacity theorem} in the data compression literature, or the regret-capacity theorem in the prediction literature. This theorem is due to~\cite{gallager1}. Its elegance lies in expressing the penalty against the optimal performance in terms of an intuitively pleasing information maximization problem.

In this paper, we study a variant of the prediction problem referred to as {\it batch universal prediction,} recently proposed in~\cite{BondaschiG:24isit, fogel2}.
This variant is motivated by the recent advent of large language models (LLMs). LLMs may be thought of as predictors. They predict relatively short sequences (which we refer to as {\it batches}), but have previously been trained on an entire corpus of such batches.
When such a process is viewed through the lens of information measures, it naturally leads to a formulation in terms of conditional information measures. The conditioning is with respect to the training corpus. In this paper, we establish a conditional regret-capacity theorem for this scenario. We show that this leads to a conditional information maximization problem. Such a result is powerful in that it allows to derive lower bounds on the minimal batch regret.

The logarithmic loss perspective has been recently generalized. In particular, logarithmic loss directly leads to the classic Kullback-Leibler divergence. A natural generalization of this divergence is R\'enyi's $\alpha$-divergence. This connection was recently leveraged to introduce the $\alpha$-regret as a generalization of the standard regret under logarithmic loss~\cite{bondaschi1}. A particularly interesting feature of this perspective is that it interpolates between the average and the worst-case logarithmic loss.
For this generalized setting, a similar regret-capacity theorem has been found~\cite{yagli1, bondaschi1}. Interestingly, the ensuing information maximization problem involves Sibson's mutual information of order $\alpha$ (see e.g.~\cite{EspositoIG:2025}).

In this paper, we also establish a regret-capacity theorem for this generalized regret measure. Not surprisingly, this involves a conditional version of Sibson's mutual information of order $\alpha.$ But it is important to note that there is no single definition of such a conditional version, see e.g. the discussion in~\cite{esposito1}.

\section{Batch Universal Prediction}

The problem of universal prediction consists in designing predictors that accurately estimates the probability distribution of the future of a sequence of symbols given the past, even when the knowledge of the underlying distribution generating the data is limited. More formally, following the setting in~\cite{BondaschiG:24isit}, given a finite input alphabet $\mathcal{X}$, for every sequence of symbols $x^i = (x_1,x_2,\dots,x_i)\in\mathcal{X}^i$, a predictor $\hat{p}$ assigns a probability $\hat{p}(y | x^{i})$ for the $(i+1)$-th symbol to be equal to $y\in\mathcal{X}$ given the past $x^{i}$. In universal prediction, a predictor is generally required to perform well if the data is generated according to \emph{any} distribution in a given parametric class $\mathcal{P} = \{p_{\theta}: \theta\in\Theta\}$ with parameter space $\Theta$. A \emph{loss function} is used to evaluate the quality of the predictor. In language models, the \emph{logarithmic} (or \emph{cross-entropy}) loss is generally employed. It is defined point-wise as 
\begin{equation}
L(\hat{p}, y | x^{i}) = -\log \hat{p}(y|x^{i}).
\end{equation}
Usually, at inference, LLMs predict a sequence of symbols in an online fashion, and the cumulative loss of the entire sequence is considered. In the case where the number of sequential symbols is $\ell,$ the cumulative loss equals 
\begin{equation}
L(\hat{p}, y^{\ell} | x^i) = -\log \hat{p}(y^{\ell} | x^i), \ \ \hat{p}(y^{\ell} | x^i) = \prod_{j=1}^{\ell} \hat{p}(y_j | x^i,y^{j-1}).
\end{equation}
For a given ground-truth distribution $p_{\theta} \in\mathcal{P}$, the \emph{regret} is defined as the difference between the loss of a candidate predictor $\hat{p}$ and that of $p_{\theta}$, i.e., 
\begin{equation}
R(\hat{p},\theta,y^l |x^i) = L(\hat{p}, y^{\ell} | x^i) - L(p_{\theta}, y^{\ell} | x^i) = \log \frac{p_{\theta}(y^{\ell})}{\hat{p}(y^{\ell} | x^i)}.
\end{equation}
The \emph{average regret} is defined as the expected regret over sequences distributed according to $p_{\theta}$,
\begin{align}
R(\hat{p},\theta) &= \mathbb{E}_p[R(\hat{p},\theta,Y^{\ell}|X^i)] \label{eq:avg-regret} \\
&= \sum_{x^i} p_{\theta}(x^i)\sum_{y^{\ell}}p_{\theta}(y^{\ell})\log \frac{p_{\theta}(y^{\ell})}{\hat{p}(y^{\ell} | x^i)}.
\end{align}
The \emph{maximal average regret} is the maximum average regret over all distributions in $\mathcal{P}$, i.e., $R(\hat{p}) = \max_{\theta\in\Theta} R(\hat{p},\theta)$.

In standard universal prediction literature, the average regret for the prediction of an entire sequence of $n$ symbols \cite{merhav1, grunwald1} is considered. More precisely, a predictor is defined to output an estimated probability for every $n$-sequence $y^n$, which is denoted by $\hat{p}(y^n)$. The average regret is then $R_n(\hat{p},\theta) = \sum_{y^n}p_{\theta}(y^n)\log \frac{p_{\theta}(y^n)}{\hat{p}(y^n)}$. This case has been studied extensively, in particular for the memoryless case, where $\mathcal{P}$ is the class of distributions generating i.i.d. symbols. For this case, the asymptotical expression for $R_n(\hat{p},\theta)$ as $n\to\infty$ has been derived \cite{xie2}. Furthermore, the \emph{add-}$\frac{1}{2}$ predictor, also called Krichevsky-Trofimov predictor, has been shown to be close to optimal asymptotically. 

In \cite{BondaschiG:24isit}, a generalization of the classical average regret, called \emph{batch regret}, is introduced, to better accommodate the idea that LLMs are trained and tested on \emph{batches} of data. In particular, during the training phase a LLM model is fed $n$ batches of data, independent of each other, each of them made of $\ell$ samples. At the end of the training phase, the LLM performance is then measured over a fresh test batch, again consisting of $\ell$ samples. In the sequel, we compactly denote each batch of training data as $x_i \in\mathcal{X}^{\ell}$, so that $x^n = (x_1, x_2, \dots, x_n)$ is a sequence of $n$ training batches, each consisting of $\ell$ samples. Similarly, the test batch is denoted by $y\in\mathcal{X}^{\ell}$. Since each batch is generated according to the ground-truth distribution independently of the others, we also denote, with abuse of notation, $p_{\theta}(x^n) = \prod_{i=1}^n p_{\theta}(x_i)$.

\begin{dfn}[Batch regret \cite{BondaschiG:24isit}]
Let $x^n = (x_1,x_2,\dots,x_n)$ be a sequence of $n$ training batches, where each $x_i \in\mathcal{X}^{\ell}$ is a sequence of $\ell$ samples. Each batch is generated independently from the others, from a certain distribution in $\mathcal{P}$. Let $\hat{p}(y | x^n)$ be a predictor that, given $x^n$, estimates the probability of a fresh batch $y \in\mathcal{X}^{\ell}$ of $\ell$ samples, generated independently of $x^n$ from the same distribution. \\ \emph{Batch regret} is then defined as
\begin{equation}
\label{eq:batch-regret}
R(\hat{p}, \theta) \triangleq \sum_{x^n} p_{\theta}(x^n) \sum_{y}p_{\theta}(y)\log \frac{p_{\theta}(y)}{\hat{p}(y | x^n)}.
\end{equation}
\end{dfn}

\section{The Conditional Regret-Capacity Theorem}
The deep connection between universal prediction under logarithmic loss and universal compression has been extensively studied in the literature. In particular, regret measures used in universal prediction can be interpreted as redundancy measures in universal compression, and therefore expressed as maximal Kullback-Leibler divergences between the distributions in the class $\mathcal{P}$ and the predictors. Leveraging on information theory results such as redundancy-capacity theorems (or regret-capacity theorems in universal prediction), maximal regret measures can be recast as minimal mutual information quantities, which in turn can be used to derive lower bounds on the optimal regret.

In the case of batch regret as defined in \eqref{eq:batch-regret}, one can naturally rewrite it as a conditional KL divergence,
\begin{equation}
R(\hat{p},\theta) = D(Y \| \hat{Y} \mid X^n).\label{eq-batchregretasKL}
\end{equation}
Then, one can prove the following regret-capacity type of result that connects the minimax batch regret to a maximal conditional mutual information.
\begin{thm}
\label{thm:cond-rc}
Let $\mathcal{X}$ be a discrete alphabet set, and let $\mathcal{P} = \{p_{\theta} : \theta\in\Theta\}$ be a parametric class of distributions on $\mathcal{X}$. Let $w$ be a prior distribution on $\Theta$. Let $\theta$ be a random variable on $\Theta$, and let $X^n, Y$ be random variables on $\mathcal{X}$ such that $(\theta, X^n, Y) \sim w(\theta) p_{\theta}(X^n) p_{\theta}(Y)$, for some prior distribution $w$ on $\Theta$. Denote by $I_w (\theta; Y \mid X^n)$ the conditional mutual information when $\theta$ is distributed according to $w$. Suppose that there exists a probability distribution $w^*$ on $\Theta$ such that
\begin{equation}
I_{w^*}(\theta;Y\mid X^n) = \sup_w I_w(\theta;Y\mid X^n).
\end{equation}
Then, the minimax batch regret is equal to
\begin{equation}
\min_{\hat{p}} \max_{\theta} R(\hat{p}, \theta) = I_{w^*}(\theta;Y\mid X^n)
\end{equation}
and the predictor that achieves the minimal regret is the conditional mixture estimator with the optimal prior distribution $w^*$,
\begin{equation}
\label{eq:p-star}
\hat{p}^*(y|x^n) = \int_{\Theta} w^*(\theta|x^n) p_{\theta}(y) d\theta
\end{equation}
where
\begin{equation}
w^*(\theta | x^n) \triangleq \frac{w^*(\theta) p_{\theta}(x^n)}{\int_{\Theta}w^*(\theta) p_{\theta}(x^n) d\theta}.
\end{equation}
\end{thm}
\begin{IEEEproof}
Let $C\triangleq \sup_w I_w(\theta;Y\mid X^n)$. We first show that the minimal regret is at least $C$, and then we prove that the regret of $\hat{p}^*(y^{\ell} |x^n)$ is at most $C$. To prove the first part, notice that
\begin{align}
\min_{\hat{p}} \max_{\theta} R(\hat{p},\theta) &= \min_{\hat{p}} \max_{\theta} D_{\theta}(Y \| \hat{Y} \mid X^n) \\
    &= \min_{\hat{p}} \max_w \mathbb{E}_w[D_{\theta}(Y \| \hat{Y} \mid X^n)] \\
    &\geq \max_w \min_{\hat{p}} \mathbb{E}_w[D_{\theta}(Y \| \hat{Y}\mid X^n)].
\end{align}
It now suffices to show that
\begin{align}
    \min_{\hat{p}} \mathbb{E}_w[D_{\theta}(Y \| \hat{Y}\mid X^n)] = I_w(\theta;Y \mid X^n).
\end{align}
Indeed, let $p(y|x^n) = \int_{\Theta} w(\theta | x^n) p_{\theta}(y) \, d\theta$. Then,
\begin{align}
\mathbb{E}_w[D_{\theta}(Y \| \hat{Y}\mid X^n)] &= \mathbb{E}_{\theta,X^n,Y}\left[\log\frac{p_{\theta}(y)}{\hat{p}(y|x^n)}\right] \\
    &= \mathbb{E}_{\theta,X^n,Y}\left[\log\frac{p_{\theta}(y)}{p(y|x^n)}\right] \notag\\
    &\hspace{3em}+\mathbb{E}_{X^n,Y}\left[\log\frac{p(y|x^n)}{\hat{p}(y|x^n)}\right] \\
    &= I_w(\theta;Y\mid X^n) + D(p\|\hat{p}\mid x^n)
\end{align}
which is minimized to $I_w(\theta;Y\mid X^n)$ by picking $\hat{p}(y|x^n) = p(y|x^n)$.

To prove the second part, it suffices to show that, for the predictor in Equation \eqref{eq:p-star}, $D_{\theta}(Y \| \hat{Y}\mid X^n) \leq C$ for every $\theta$. We prove this by contradiction. Suppose that there exists a $\tilde{\theta}\in\Theta$ such that $D_{\tilde{\theta}}(Y \| \hat{Y} \mid X^n) > C$. Take 
\begin{equation}
\tilde{w}_t(\theta) = (1-t)w^*(\theta) + t\delta_{\tilde{\theta}}
\end{equation}
as the prior distribution on $\Theta$. Since $w^*$ minimizes $I_w(\theta,Y\mid X^n)$, the derivative of the function
\begin{equation}
f(t) = I_{\tilde{w}_t}(\theta;Y \mid X^n)
\end{equation}
must be non-negative at $t=0$. However, one can easily check that
\begin{equation}
f'(0) = D_{\tilde{\theta}}(Y \| \hat{Y}\mid X^n) - C > 0
\end{equation}
since we assumed $D_{\tilde{\theta}}(Y \| \hat{Y}\mid X^n) > C$. This contradicts the fact that $w^*$ minimizes $I_w(\theta;Y\mid X^n)$, and the second part of the theorem is proved.
\end{IEEEproof}

\section{Lower Bound on the Batch Regret for Binary IID Sources}
The usefulness of the conditional Regret-Capacity theorem is that it can be used to derive lower bounds on the minimal batch regret. In this section, we exemplify this by applying Theorem \ref{thm:cond-rc} to the case of binary i.i.d. sources. More formally, let $\mathcal{X}=\{0,1\}$ be the binary alphabet, and let $\mathcal{P}_{\rm iid}$ be the class of memoryless sources, i.e., sources generating i.i.d. binary digits with a given probability,
\begin{equation}
\label{eq:iid-sources}
\mathcal{P}_{\rm iid} = \{p_{\theta}(x^i) = \theta^{n_1} (1-\theta)^{n_0}, \theta \in [0,1], \text{ for any }i\in\mathbb{N}^+\}
\end{equation}
where $n_1$ and $n_0$ are the number of ones and zeros in $x^i$, respectively. In this setting, the sequence of $n$ batches $X^n$, each consisting of $\ell$ samples, is equivalent to one single batch on length $n\ell$, due to the iid nature of the data. Hence, for this case the average batch regret is
\begin{multline}
R(\hat{p},\theta) = \sum_{t_1 = 0}^{t} \binom{t}{t_1}\theta^{t_1} (1-\theta)^{t_0} \\
\sum_{\ell_1=0}^{\ell} \binom{\ell}{{\ell}_1} \theta^{{\ell}_1} (1-\theta)^{{\ell}_0} \log \frac{\theta^{\ell_1} (1-\theta)^{\ell_0}}{\hat{p}(y^{\ell} | x^t)}
\end{multline}
where the number of zeroes and ones in $X^n$ and $Y,$ respectively, are denoted as $t_0$, $t_1$, $\ell_0$, and $\ell_1,$ and we have 
$t=n\ell.$

In \cite{BondaschiG:24isit}, the authors study the batch regret for the conditional add-constant predictor, which is defined as
\begin{equation}
\label{eq:pred-def}
\hat{p}_{\beta}(y^{\ell} | \bm{x}^n) = \prod_{i=1}^{\ell} \hat{p}_{\beta}(y_i | \bm{x}^n, y^{i-1}),
\end{equation}
where
\begin{equation}
\label{eq:add-beta}
\hat{p}_{\beta}(y_i=1 | \bm{x}^n, y^{i-1}) = \frac{t_1 + \ell_1^{(i-1)} + \beta}{t + i - 1 + 2\beta}
\end{equation}
for a chosen $\frac{1}{2} \leq \beta \leq 1$. This predictor is actually a conditional mixture predictor, with the prior $w$ taken as the symmetric Dirichlet distribution with parameter $\beta$, i.e.,
\begin{equation}
w(\theta) = \frac{\Gamma(2\beta)}{\Gamma^2(\beta)} \theta^{\beta-1}(1-\theta)^{\beta-1}.
\end{equation}
For this predictor, \cite[Theorem 1]{BondaschiG:24isit} shows that the batch regret in the interval $\Theta_{\delta} = [\delta, 1-\delta]$, for any $0<\delta<\frac{1}{2}$, is equal to
\begin{equation}
\max_{\theta \in\Theta_{\delta}} R(\hat{p}_{\beta},\theta) = \frac{1}{2}\log \left(1+\frac{1}{n}\right) + O\left(\frac{1}{n\ell}\right).
\end{equation}
Using Theorem \ref{thm:cond-rc}, we can show that the minimal batch regret is asymptotically the same, essentially showing that all conditional add-constant estimators are asymptotically optimal in $\Theta_{\delta}$, provided that the appropriate regime for $n$ and $\ell=\ell(n)$ is considered.

\begin{thm}
For the class of i.i.d. distributions $\mathcal{P}_{\rm iid}$, if ${\ell = \Theta(n^{\gamma})}$ for $\gamma > 0$, the following lower bound on the minimal batch regret holds,
\begin{equation}
\min_{\hat{p}} \max_{\theta\in[0,1]} R(\hat{p},\theta) \geq \frac{1}{2} \log \left(1 + \frac{1}{n}\right) + O\left(\frac{\log(n\ell)}{n\ell}\right).
\end{equation}
\end{thm}
\begin{IEEEproof}
Let $w(\theta)$ be the Dirichlet distribution with parameter $1$, that is, the uniform distribution on $\Theta=[0,1]$. Then, by Theorem \ref{thm:cond-rc}, 
\begin{align}
\min_{\hat{p}} \max_{\theta} R(\hat{p},\theta) &\geq I_w(\theta, Y\mid X^n) \\
&= \mathbb{E}_{\theta \sim w} \big[D_{\theta}(Y\|\hat{Y}\mid X^n)\big] \\
&= \mathbb{E}_{\theta \sim w} \left[R(\hat{p}_w, \theta)\right] \label{eq:exp-regret}
\end{align}
where 
\begin{equation}
\hat{p}_w(y^z) = \int_0^1 w(\theta) p_{\theta}(y^z) d\theta.
\end{equation}
By inspecting the proof of \cite[Theorem 1]{BondaschiG:24isit}, which generalizes the proof of \cite[Proposition 1]{xie2} for $\beta > \frac{1}{2}$, we get that
\begin{equation}
R(\hat{p}_w,\theta) \geq \frac{1}{2}\log\left(1+\frac{1}{n}\right) - \frac{5}{n\ell} - \frac{5}{n\ell}\cdot\frac{1}{\theta(1-\theta)}
\end{equation}
for every $\theta \in \left[\frac{1}{n\ell}, 1-\frac{1}{n\ell}\right]$.

Hence, we have that
\begin{align}
\min_{\hat{p}} &\max_{\theta} R(\hat{p},\theta) - \frac{1}{2}\log\left(1+\frac{1}{n}\right) \\
&\geq \int_{1/n\ell}^{1-1/n\ell} R(\hat{p}_w,\theta)\,d\theta - \frac{1}{2}\log\left(1+\frac{1}{n}\right) \\
&= \int_{1/n\ell}^{1-1/n\ell} \left(R(\hat{p}_w,\theta)- \frac{1}{2}\log\left(1+\frac{1}{n}\right)\right)d\theta \\
&\hspace{13em}- \log\left(1+\frac{1}{n}\right)\cdot \frac{1}{n\ell}\\
&\geq -\frac{5}{n\ell}\int_{1/n\ell}^{1-1/n\ell} \frac{1}{\theta(1-\theta)}d\theta - \frac{5}{n\ell} - \frac{1}{n^2\ell}.
\end{align}
Now note that
\begin{align}
\int_{1/n\ell}^{1-1/n\ell} \frac{1}{\theta(1-\theta)}d\theta &= 2 \int_{1/n\ell}^{1/2} \frac{1}{\theta(1-\theta)}d\theta \\
&\leq 4\int_{1/n\ell}^{1/2} \frac{1}{\theta}\,d\theta \\
&= 4\log(n\ell) - 4\log 2,
\end{align}
and therefore,
\begin{align}
\min_{\hat{p}} \max_{\theta} R(\hat{p},\theta) - &\frac{1}{2}\log\left(1+\frac{1}{n}\right) \\
&\geq -\frac{20\log(n\ell)}{n\ell} + \frac{9}{n\ell} - \frac{1}{n^2\ell}.
\end{align}
\end{IEEEproof}

\section{Extension to R\'enyi and Sibson Information Measures}
A (non-batched) regret measure based on the R\'{e}nyi divergence, called $\alpha$-regret, was introduced in \cite{bondaschi1}: for $\alpha \geq 1$,
\begin{align}
R_{\alpha}^{\rm NB}(\hat{p},\theta) &= D_{\alpha}(Y \| \hat{Y}) \\
&= \frac{1}{\alpha - 1}\log\left\{\sum_{y} p_{\theta}(y) \left(\frac{p_{\theta}(y)}{\hat{p}(y)}\right)^{\alpha -1}\right\}.
\end{align}
The measure can be operationally interpreted as an intermediate between average and worst-case regrets. Specifically, for $\alpha=1,$ we recover the standard average regret, and in the limit $\alpha\rightarrow\infty,$ the classic worst-case regret. For this regret measure, a regret-capacity theorem was derived in~\cite[Theorem 1]{bondaschi1}, involving Sibson's mutual information of order $\alpha$ in place of the regular mutual information appearing in the classic regret-capacity theorem.

The $\alpha$-regret measure can be naturally extended to the batched case using the conditional R\'{e}nyi divergence in place of its unconditional counterpart. That is, Equation~\eqref{eq-batchregretasKL} is generalized to
\begin{align}
\lefteqn{R_{\alpha}(\hat{p},\theta)} \nonumber \\
&= D_{\alpha}(Y \| \hat{Y} | X^n) \\
&= \frac{1}{\alpha - 1}\log\left\{\sum_{x^n} p_{\theta}(x^n)\sum_{y} p_{\theta}(y) \left(\frac{p_{\theta}(y)}{\hat{p}(y|x^n)}\right)^{\alpha -1}\right\}
\end{align}
As pointed out, this measure can also be interpreted as a middle way between the two classic regret measures. Formally, in the limit as $\alpha \to 1$, one retrieves the average batch regret defined above, namely,
$\lim_{\alpha \to 1} R_{\alpha}(\hat{p},\theta) = R(\hat{p},\theta).$
In the limit $\alpha \to \infty$, one gets the worst-case batch regret, which is in turn equivalent to the non-batched one:
\begin{align}
\lim_{\alpha\to\infty} R_{\alpha}(\hat{p},\theta) &= \max_{x^n} \max_y \log \frac{p_{\theta}(y)}{\hat{p}(y|x^n)} \\
&= \max_y \log \frac{p_{\theta}(y)}{\hat{p}(y)}.
\end{align}

A type of conditional regret-capacity theorem can be derived also for this regret measure, where taking the place of the classical conditional mutual information is another information measure, which was proposed in the literature as a possible conditional version, among others, of Sibson's mutual information \cite{tomamichel1, esposito1}.
\begin{dfn}
Let $\theta, X, Y, \hat{Y}$ be random variables such that $(\theta, Y) - X - \hat{Y}$, and $Y$ and $\hat{Y}$ are defined on the same alphabet. Let $\hat{p}$ be the conditional distribution of $\hat{Y}$ given $X$. Then, the conditional Sibson's mutual information is defined as
\begin{equation}
\label{eq:cond-sibson}
I_{\alpha}(\theta, Y \mid X) \triangleq \min_{\hat{p}} D_{\alpha}(Y \| \hat{Y} \mid \theta, X).
\end{equation}
\end{dfn}
For the case considered in this paper, the following useful property can be proved, by minimizing each conditional distribution $\hat{p}(y|x^n)$ separately.
\begin{thm}
Let $\mathcal{X}$ be a discrete alphabet set, and let $\mathcal{P} = \{ p_{\theta}: \theta\in\Theta\}$ be a parametric class of distributions on $\mathcal{X}$. Let $\theta$ be a random variable on $\Theta$, and let $X^n, Y$ be random variables on $\mathcal{X}$ such that $(\theta, X^n, Y) \sim w(\theta) p_{\theta}(X^n) p_{\theta}(Y)$, for some prior distribution $w$ on $\Theta$, where with abuse of notation, we denote $p_{\theta}(x^n) = \prod_{i=1}^n p_{\theta}(x_i)$. Then, the conditional Sibson's mutual information is equal to
\begin{multline}
\label{eq:cond-sibson2}
I_{\alpha}(\theta, Y \mid X^n) \\
= \frac{1}{\alpha-1} \log \sum_{x^n} \left\{\sum_y\left(\int_{\Theta} w(\theta)p_{\theta}(x^n) p_{\theta}(y)^{\alpha}d\theta\right)^{1/\alpha}\right\}^{\alpha}
\end{multline}
and the minimizing distribution in Equation \eqref{eq:cond-sibson} is
\begin{equation}
\label{eq:opt-alpha}
\hat{p}_{\alpha}(y|x^n) \triangleq \frac{\left\{\int_{\Theta} w(\theta|x^n) p_{\theta}(y)^{\alpha} d\theta\right\}^{1/\alpha}}{\sum_y \left\{\int_{\Theta} w(\theta|x^n) p_{\theta}(y)^{\alpha} d\theta\right\}^{1/\alpha}}
\end{equation}
where
\begin{equation}
w(\theta | x^n) \triangleq \frac{w(\theta) p_{\theta}(x^n)}{\int_{\Theta}w(\theta) p_{\theta}(x^n) d\theta}.
\end{equation}
\end{thm}
The closed-form formula in Equation \eqref{eq:cond-sibson2} is a special case of the formula derived in \cite{tomamichel1} when $X^n - \theta - Y$.
The predictor in Equation \eqref{eq:opt-alpha} is a conditional version of the $\alpha$-NML predictor introduced in \cite{bondaschi1}.

The regret-capacity theorem for the batch $\alpha$-regret takes the following form.
\begin{thm}
Let $\mathcal{X}$ be a discrete alphabet set, and let $\mathcal{P} = \{p_{\theta} : \theta\in\Theta\}$ be a parametric class of distributions on $\mathcal{X}$. Let $w$ be a prior distribution on $\Theta$. Let $\theta$ be a random variable on $\Theta$, and let $X^n, Y$ be random variables on $\mathcal{X}$ such that $(\theta, X^n, Y) \sim w(\theta) p_{\theta}(X^n) p_{\theta}(Y)$, for some prior distribution $w$ on $\Theta$. 
Denote the conditional Sibson's mutual information when $\theta$ is distributed according to $w$ by $I_{\alpha}^w (\theta, Y \mid X^n)$ .
Suppose that there exists a probability distribution $w^*$ on $\Theta$ such that
\begin{equation}
I_{\alpha}^{w^*}(\theta,Y\mid X^n) = \sup_w I_{\alpha}^w(\theta,Y\mid X^n).
\end{equation}
Then, the minimax batch $\alpha$-regret is equal to
\begin{equation}
\min_{\hat{p}} \max_{\theta} R_{\alpha}(\hat{p}, \theta) = I_{\alpha}^{w^*}(\theta,Y\mid X^n)
\end{equation}
and the predictor that achieves the minimal regret is the conditional $\alpha$-NML with the optimal prior distribution $w^*$,
\begin{equation}
\label{eq:p-alpha-star}
\hat{p}_{\alpha}^*(y|x^n) = \frac{\left\{\int_{\Theta} w^*(\theta|x^n) p_{\theta}(y)^{\alpha} d\theta\right\}^{1/\alpha}}{\sum_y \left\{\int_{\Theta} w^*(\theta|x^n) p_{\theta}(y)^{\alpha} d\theta\right\}^{1/\alpha}}
\end{equation}
where
\begin{equation}
w^*(\theta | x^n) \triangleq \frac{w^*(\theta) p_{\theta}(x^n)}{\int_{\Theta}w^*(\theta) p_{\theta}(x^n) d\theta}.
\end{equation}
\end{thm}
\begin{IEEEproof}
The proof is similar to the average batch regret case. Let $C_{\alpha}\triangleq \sup_w I_{\alpha}^w(\theta,Y\mid X^n)$. We first show that the minimal batch $\alpha$-regret is at least $C_{\alpha}$. To that end, notice that
\begin{align}
\min_{\hat{p}} \max_{\theta} R(\hat{p},\theta) &= \min_{\hat{p}} \max_{\theta} \frac{1}{\alpha-1}\log \mathbb{E}_{X^n, Y \sim p_{\theta}}\left[\frac{p_{\theta}(Y)}{\hat{p}(Y|X^n)}\right] \\
&= \min_{\hat{p}} \frac{1}{\alpha-1}\log \max_{\theta}\mathbb{E}_{X^n, Y \sim p_{\theta}}\left[\frac{p_{\theta}(Y)}{\hat{p}(Y|X^n)}\right] \\
&= \min_{\hat{p}} \frac{1}{\alpha-1}\log \max_w \mathbb{E}_{\theta, X^n, Y}\left[\frac{p_{\theta}(Y)}{\hat{p}(Y|X^n)}\right] \\
&\geq \max_w \min_{\hat{p}} D_{\alpha}(Y \| \hat{Y} \mid \theta, X^n) \\
&= \max_w I_{\alpha}(\theta, Y \mid X^n) = C_{\alpha}.
\end{align}

We now show that for the conditional $\alpha$-NML in Equation \eqref{eq:p-alpha-star}, for every $\theta\in\Theta$, for $(X^n, Y, \hat{Y}) \sim p_{\theta}(X^n) p_{\theta}(Y) \hat{p}^*_{\alpha}(Y|X^n)$, we have $D_{\alpha}(Y \| \hat{Y}\mid X^n) \leq C_{\alpha}$. Suppose by contradiction that there exists a $\tilde{\theta}\in\Theta$ such that $D_{\alpha}(Y \| \hat{Y} \mid X^n) > C_{\alpha}$. Take 
\begin{equation}
w_t(\theta) = (1-t)w^*(\theta) + t\delta_{\tilde{\theta}}
\end{equation}
as the prior distribution on $\Theta$. Since $w^*$ minimizes $I_{\alpha}^{w}(\theta,Y\mid X^n)$, the derivative of the function
\begin{equation}
f_{\alpha}(t) = I_{\alpha}^{w_t}(\theta,Y \mid X^n)
\end{equation}
must be non-negative at $t=0$. However, notice that we have
\begin{multline}
f_{\alpha}(t) = \frac{1}{\alpha-1}\log \sum_{x^n} \Big\{\sum_{y}\Big(tp_{\tilde{\theta}}(x^n)p_{\tilde{\theta}}(y)^{\alpha} \\
+ (1-t)\sum_{\theta}w^*(\theta)p_{\theta}(x^n)p_{\theta}(y)^{\alpha}\Big)^{1/\alpha}\Big\}^{\alpha}.
\end{multline}
Its derivative is
\begin{multline}
f'_{\alpha}(t) = \\
\frac{1}{Z_{\alpha}(t)} \sum_{x^n}\Big\{\sum_y \Big(\sum_{\theta} w_t(\theta) p_{\theta}(x^n)p_{\theta}(y)^{\alpha}\Big)^{1/\alpha}\Big\}^{\alpha-1} \\
\cdot \sum_y\Big(\sum_{\theta} w_t(\theta)p_{\theta}(x^n)p_{\theta}(y)^{\alpha}\Big)^{\frac{1-\alpha}{\alpha}} \\
\cdot \Big(p_{\tilde{\theta}}(x^n)p_{\tilde{\theta}}(y)^{\alpha} - \sum_{\theta} w^*(\theta) p_{\theta}(x^n)p_{\theta}(y)^{\alpha}\Big)
\end{multline}
where
\begin{equation}
Z_{\alpha}(t) = (\alpha-1) \exp\left\{(\alpha-1)f_{\alpha}(t)\right\}.
\end{equation}
Hence, at $t=0$, we have
\begin{equation}
f'_{\alpha}(0) = \frac{\exp\{(\alpha-1)D_{\alpha}(Y\| \hat{Y}\mid X^n)\}}{\exp\{(\alpha-1)I_{\alpha}^{w^*}(\theta,Y \mid X^n)\}} -1.
\end{equation}
Since by assumption $D_{\alpha}(Y \| \hat{Y} \mid X^n) > C_{\alpha}$, we have that $f'(0)>0$, which is a contradiction. Hence, we must have $D_{\alpha}(Y \| \hat{Y} \mid X^n) \leq C_{\alpha}$ for every $\tilde{\theta}$, and therefore $\max_{\theta} R_{\alpha}(\hat{p}^*_{\alpha},\theta) \leq C_{\alpha}$, concluding the proof.
\end{IEEEproof}

\section*{Acknowledgements}

The work in this paper was supported in part by the Swiss National Science Foundation under Grant 200364.

\bibliographystyle{IEEEtran}
\bibliography{prediction}

\end{document}